\documentclass[sigconf]{acmart}

\AtBeginDocument{%
  \providecommand\BibTeX{{%
    \normalfont B\kern-0.5em{\scshape i\kern-0.25em b}\kern-0.8em\TeX}}}

\copyrightyear{2020} 
\acmYear{2020} 
\setcopyright{acmlicensed}\acmConference[CHI PLAY '20]{Proceedings of the Annual Symposium on Computer-Human Interaction in Play}{November 2--4, 2020}{Virtual Event, Canada}
\acmPrice{15.00}
\acmDOI{10.1145/3410404.3414247}
\acmISBN{978-1-4503-8074-4/20/11}

\newcommand{\FG}[1]{Figure~\ref{#1}}
\newcommand{\TA}[1]{Table~\ref{#1}}
\newcommand{\EQ}[1]{Equation~\ref{#1}}

\newcommand{\SC}[1]{Section~\ref{#1}}

\usepackage{microtype}  
\usepackage{amsmath}
\usepackage{esvect}
\usepackage{enumitem}
\usepackage{booktabs}
\usepackage{url}
\usepackage{soul}

\begin{document}

\begin{teaserfigure}
  \includegraphics[width=\textwidth]{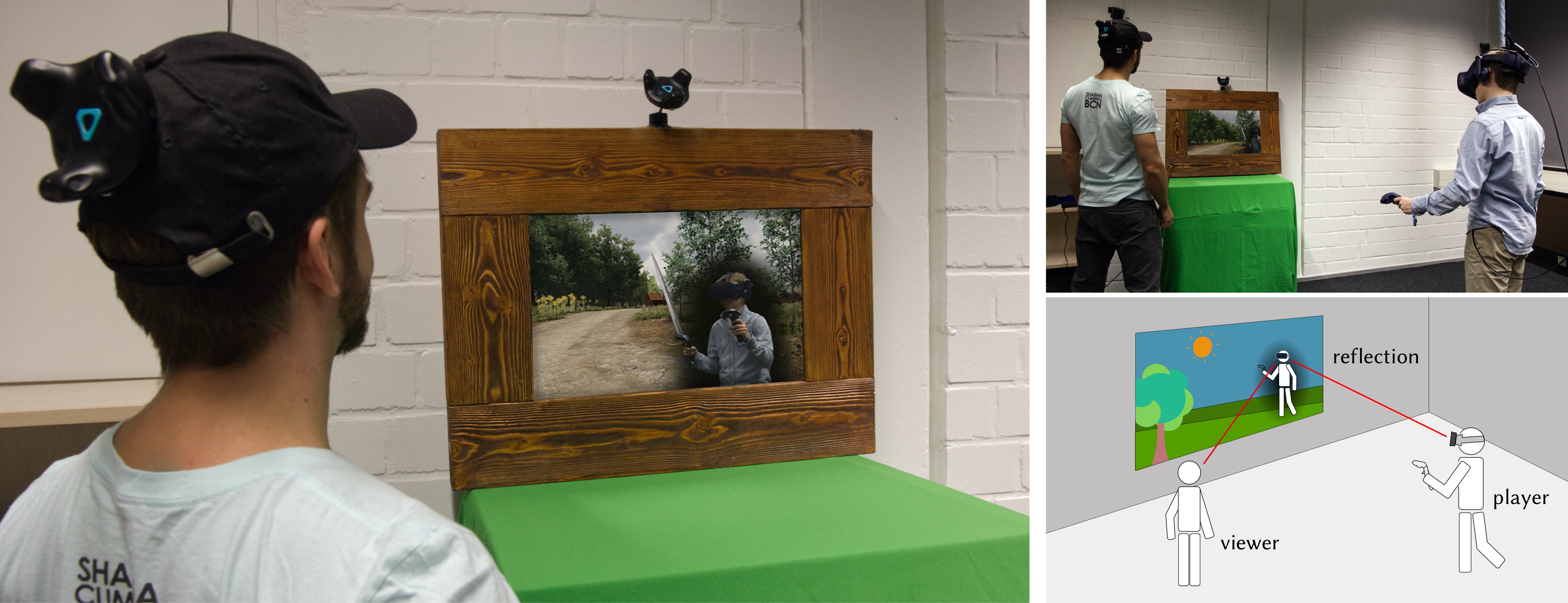}
  \caption{Our paper contributes a new spectatorship experience for virtual reality games. We blend the player's reflection in the virtual environment and provide a dynamic view frustum that allows viewers to explore the game world by themselves.}
  \label{fig:teaser}
\end{teaserfigure}

\title[Silhouette Games: An Interactive One-Way Mirror Approach]{Silhouette Games: An Interactive One-Way Mirror Approach to Watching Players in VR}

\author{Andrey Krekhov}
\affiliation{%
  \institution{University of Duisburg-Essen}
  \city{Duisburg}
  \state{Germany}
}
\email{andrey.krekhov@uni-due.de}

\author{Daniel Preu{\ss}}
\affiliation{%
  \institution{University of Duisburg-Essen}
  \city{Duisburg}
  \state{Germany}
}
\email{daniel.preuss@uni-due.de}

\author{Sebastian Cmentowski}
\affiliation{%
  \institution{University of Duisburg-Essen}
  \city{Duisburg}
  \state{Germany}
}
\email{sebastian.cmentowski@uni-due.de}

\author{Jens Kr{\"u}ger}
\affiliation{%
  \institution{University of Duisburg-Essen}
  \city{Duisburg}
  \state{Germany}
}
\email{jens.krueger@uni-due.de}

\renewcommand{\shortauthors}{Krekhov et al.}

\begin{abstract}

Watching others play is a key ingredient of digital games and an important aspect of games user research. However, spectatorship is not very popular in virtual reality, as such games strongly rely on one's feelings of presence. In other words, the head-mounted display creates a barrier between the player and the audience. We contribute an alternative watching approach consisting of two major components: a dynamic view frustum that renders the game scene from the current spectator position and a one-way mirror in front of the screen. This mirror, together with our silhouetting algorithm, allows seeing the player's reflection at the correct position in the virtual world. An exploratory survey emphasizes the overall positive experience of the viewers in our setup. In particular, the participants enjoyed their ability to explore the virtual surrounding via physical repositioning and to observe the blended player during object manipulations. Apart from requesting a larger screen, the participants expressed a strong need to interact with the player. Consequently, we suggest utilizing our technology as a foundation for novel playful experiences with the overarching goal to transform the passive spectator into a collocated player.

\end{abstract}



\maketitle

\section{Introduction}


Looking over the player's shoulder has always been a crucial part of our digital gaming experience. There are manifold reasons to watch others play, such as entertainment or learning game strategies~\cite{sjoblom2017people}. Throughout history, the habit of observing players has evolved from casual get-togethers in domestic settings to highly attended live tournaments with thousands of on-site supporters and even more remote viewers. Not only the competitive scene has evolved: nowadays, many players share their daily gameplay sessions via live streaming platforms like Twitch~\cite{twitch}.

In virtual reality (VR), the head-mounted display (HMD) imposes an artificial barrier between viewers and the player, and enjoying the virtual world often remains a lonely experience. Although current technologies allow us to see the virtual world through the player's eyes via local or remote live streaming, the unique selling point of VR---namely the induced presence---remains unavailable to the audience.

What we see at best are the player's movements in the real-world next to a screen with the corresponding first-person view. It remains up to our imagination to merge these two perspectives and to project the player into the virtual environment. Indeed, we can automate this projection by utilizing a green screen setup~\cite{LIV}, in which case the audience can see the captured player at the correct virtual position and scale. However, due to the inherent complexity and expensiveness, such setups remain limited to high fidelity video productions required for game teasers and advertising. No less important issue of a green screen is the predefined camera position. If something is happening right outside of the view frustum, the audience will inevitably miss this event. 

To summarize, observing someone playing in VR remains an unsolved problem: On one hand, the first-person view fails to convey the full-body interaction, which is a crucial component of player experience in VR. On the other hand, green screen setups, apart from being expensive and requiring lots of effort, work only for a handful of games, such as Beat Saber\cite{BeatSaber}, where we can predict the optimal camera orientation for the whole game session. 

This paper contributes to this topic by presenting a novel possibility to watch players in VR. Our interactive approach displays the player in the virtual environment similar to a green screen caption, yet allowing the viewers to control the frustum via physical repositioning (cf. \FG{fig:tracking}). In this way, viewers perceive the visible content as a window into the virtual world rather than a static display---changing position relative to the screen offers different perspectives and prevents missing out on exciting events.

The core idea revolves around the principle of one-sided mirrors. Behind such a mirror, we place a screen that displays the virtual world depending on the observer's point of view. Our silhouetting algorithm renders a dark overlay at the position of the player's reflection to trigger the mirroring property of this local area. As a result, the observer sees the player's mirror image ``inside'' the virtual world, as shown in \FG{fig:teaser}. 

\begin{figure}[t!]
\centering
\includegraphics[width=1.0\columnwidth]{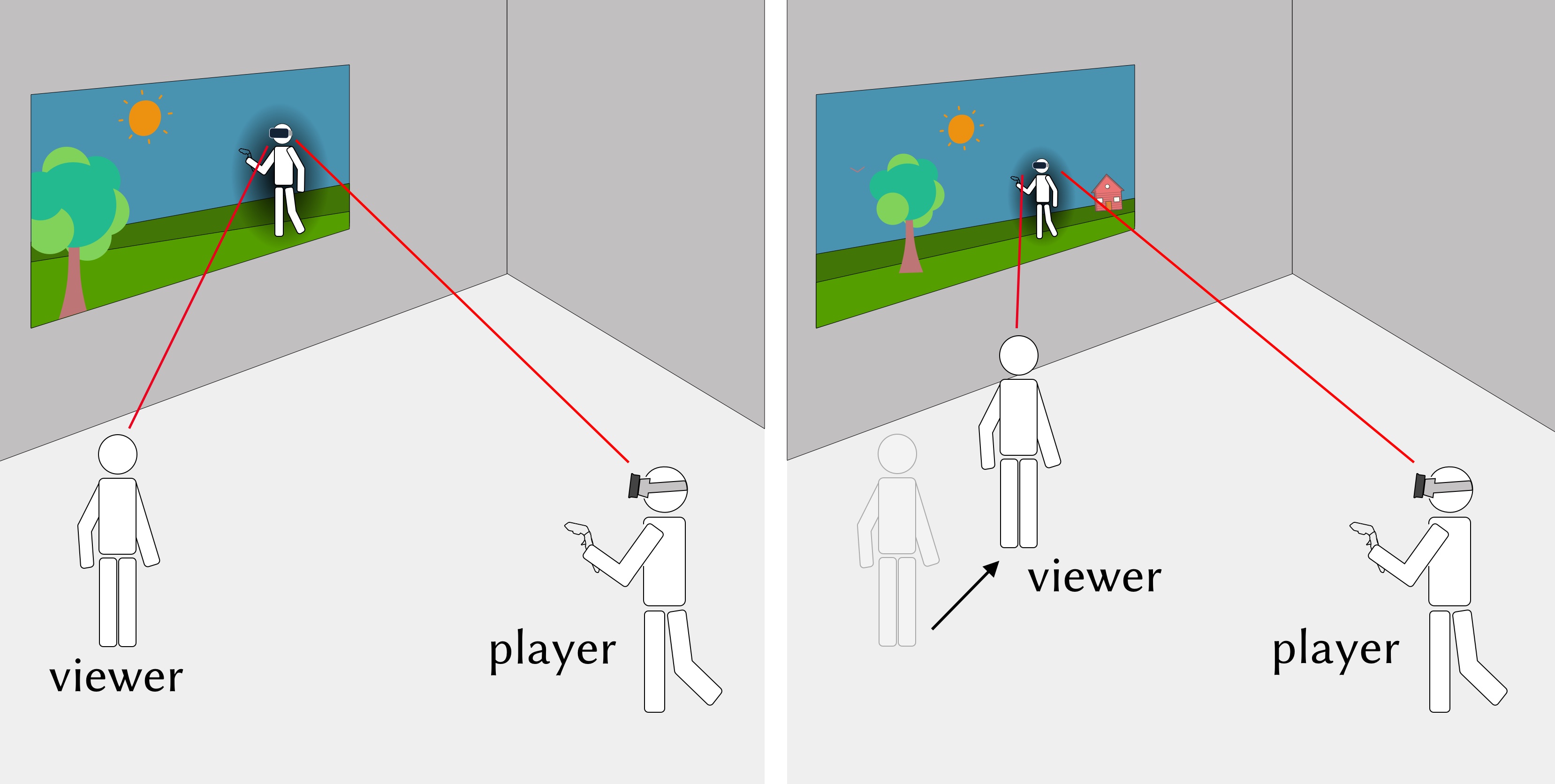}
\caption{The view frustum is rendered based on the viewer's and player's positions. The viewer, equipped with an HTC Vive Tracker, moves freely in the room to obtain the desired perspective on the virtual environment. In the example above, the viewer moves forward and to the right. This transition increases the field of view, and, e.g., reveals a house in the right corner.}
\label{fig:tracking}
\end{figure}

To gather first impressions of this alternative watching experience, we conducted a qualitative evaluation using a VR adventure game as a testbed. The outcomes reveal particular strengths of the idea, such as the overall comprehensibility of the player's actions and the highly appreciated freedom of autonomous exploration. The collected requests of the participants provide a ground for follow-up research and outline possible application areas of our collocated setup.

\section{Related Work}

Games have always attracted a broad watching audience. In the early days, video games were mostly limited to entertainment machines installed in amusement arcades and surrounded by crowds of interested bystanders~\cite{newman2013videogames}. Today, spectating is no longer bound to physical presence. Instead, viewers from all over the world can watch their favorite players using video and streaming platforms, e.g., YouTube~\cite{youtube} or Twitch~\cite{twitch}.

Nowadays, spectator is not just a passive onlooker. For instance, the game \textit{Drawful 2}~\cite{drawful} requires the spectators to guess the original phrase based on the player's hand-drawn image. This trend also extends to VR, rendering the virtual experience more social and inclusive. One recent game is \textit{Acron: Attack of the Squirrels!}~\cite{acron}. The non-VR spectators (or players) act as a team of squirrels and attempt to steal the ``Golden Acrons'', which are protected by the VR player. In contrast, the two-player version of \textit{Carly and the Reaperman: Escape from the Underworld}~\cite{carly} requires a close co-operation between the VR player and the non-VR participant in order to overcome the in-game obstacles.

Throughout the years, spectating games has been of ongoing interest to the research community~\cite{cheung2011starcraft, drucker2003spectator, frome2004helpless, sjoblom2017people, smith2013live}. Especially the recording and streaming of gameplay sessions using online platforms has recently attracted attention due to its importance for the growing esports community~\cite{carlsson2015designing, hamari2017esports, rambusch2017pre, seo2016beyond}. Apart from these distributed approaches using online media, a growing research corpus has focused on the local spectator experience. In this area, the differentiation between active player and passive audience~\cite{taylor2016play} is usually not applicable. Instead, the recent work broadens the definition of spectators to active bystanders interacting with various interfaces~\cite{tekin2017ways}, often in the form of interactive displays~\cite{muller2012looking, o2013naturalness, ten2012chained}.

Common assumptions tend to present gaming as a mostly solitary activity being shared among the active players only. Contrary to this popular belief, social activities, including spectating, are essential components of play~\cite{voida2009wii}. People like to watch others play games for various reasons, such as being entertained, distracted, or learning new game strategies~\cite{sjoblom2017people}. These desires require a sufficient understanding of the course of action and the players' activities. Reeves et al.~\cite{reeves2005designing} classify spectator interfaces based on this information being given to or hidden from the audience. In particular, the authors differentiate between the manipulations conducted by the players and the effects in the game.
Based on these two variables, the authors derive four design strategies.
For instance, a \textit{secretive} strategy is achieved if both the effects and the manipulations remain hidden. Revealing the effects but hiding the manipulations is more suited for a \textit{magical} interaction. Thus, most gaming activities are better understood and watched with a high information level on both - manipulations and effects.

In contrast to watching, for example, pre-recorded videos, local spectatorship is usually a highly interactive process. Assisting the players is commonly referred to as collocated gaming~\cite{cox2016public, kappen2014engaged, wehbe2015towards}. The grade of spectator participation can range from entirely passive viewing to actively aiding the players in their task~\cite{maurer2015gaze}. Downs et al.~\cite{downs2015differentiated} propose three durable roles: players, audience members, and bystanders. However, the audience is not a static entity and consists of spontaneous and rather ephemeral roles. Such roles include, e.g., commentators, directors, coaches, and cheerleaders.

Another example of viewer interaction is responsive displays. These interfaces can provide unique opportunities for the audience to interact with the game, e.g., by altering their view on the game or assisting the player. In their work, Maurer et al.~\cite{maurer2015gaze} evaluated how spectators could assist or interfere with the gameplay using gaze tracking.

When using such responsive displays, most users start interacting in a playful manner~\cite{tomitsch2014cares}. Instead of exploring the displayed content, the viewers usually begin to experiment with the available controls and play with their natural gestures. After satisfying their curiosity, they will ultimately shift their focus to the content. This behavior demonstrates the importance of supporting spontaneous and natural interactions as part of the spectator experience. Designing responsive displays in such a user-centric manner can also foster the well-known honeypot effect~\cite{wouters2016uncovering}.


All of the previously covered aspects of spectatorship apply to most games in general. However, the particular case of virtual reality (VR) games introduces different and challenging conditions: Players wear head-mounted displays (HMDs) to immerse fully into a virtual world. These devices replace the traditional monitor and hide the gameplay from the potential audience. At the same time, VR games often rely heavily on full-body interactions that are readily observable by bystanders. Therefore, typical VR devices can be classified as "suspenseful" interfaces according to the taxonomy by Reeves et al.~\cite{reeves2005designing}: Manipulations are revealed, while effects tend to stay hidden. These characteristics of VR setups do not provide sufficient insights for potential spectators. Another limitation is the social acceptability of HMDs, including AR and VR devices~\cite{alallah2018performer, denning2014situ, koelle2015don}. A spectator might be suspicious of being video-recorded, which results in an underwhelming experience. Therefore, it is crucial to provide the audience with a maximum amount of possible information. Many VR games offer a mirrored view of the players' sight, allowing bystanders to watch the action on a nearby display. However, this perspective does not display the player's full-body actions in the virtual environment, resulting in a loss of information.

Another issue of the HMD-induced barrier is that the player cannot see the audience. Nevertheless, the players are usually still aware that onlookers are watching their full-body movements. This self-consciousness might cause discomfort and induce the feeling of insecurity~\cite{10.1145/3290605.3300644}. On the other hand, such uncomfortable or even embarrassing interactions~\cite{mitchell2015embarrassing} also have potential benefits. Examples for this are entertainment and sociality, as discussed by Benford et al.~\cite{10.1145/2207676.2208347}.

Other approaches provide the spectators with headsets themselves to achieve a shared technological basis. For instance, Larsson et al.~\cite{larsson2001actor} evaluate the scenario of recorded VR presentations being watched from the actor's perspective using HMDs. Sra et al.~\cite{10.1145/2984751.2984779} emphasize the importance of equal experiences for all participants in a shared VR space, even if the physical size and shape of the users' individual space is significantly different. In particular, the authors consider three different spatial mapping approaches to allow everyone to perform locomotion based on physical walking~\cite{10.1145/3196709.3196788}. Research on such topics, commonly referred to as collaborative virtual environments (CVE)~\cite{benford1997crowded, benford2001collaborative, churchill2012collaborative}, mostly focuses on distributed~\cite{broll1995interacting} or asymmetric~\cite{welsfordAsymmetric} spectator experiences for specific purposes, e.g., visualizations~\cite{chastine2005ammp} or education~\cite{pan2006virtual}. In our work, we specifically target the case of non-VR viewers watching VR players.

Instead of moving the observers into the virtual reality, some recent work has blended the elements from the virtual world into the real environment. Zappi et al.~\cite{zappi2011design} used stereoscopic shutter glasses to present interactive virtual objects and real actors on the same stage and achieve a hybrid-reality performance. Jonas et al.'s \textit{RoomAlive}~\cite{jones2014roomalive} went one step further and turned the whole room into a mixed reality environment shared among players and observers using projector-camera systems. In their work, Hartmann et al.~\cite{hartmannRealityCheck} used Kinect cameras to blend the real surrounding into the virtual environment. Similarly, they also integrated a spectator mode projecting the virtual world onto the surrounding walls.

The third type of spectator interface extends the basic approach of showing the players' view on a display. Software solutions such as LIV~\cite{LIV} use green screens and external cameras to blend the real player into a third-person view of the game environment. The result provides a better impression of the players' experience. Gugenheimer et al.~\cite{gugenheimer2017sharevr} take this metaphor one step further. Their \textit{ShareVR} prototype uses a tracked handheld display to achieve a mobile window into the virtual reality that can be moved freely by the spectator. This approach is closely related to our work. With our idea, we replace the handheld device with a very intuitive mirror implementation that also achieves a blending of player and environment.

\begin{figure*}[t]
\centering
\includegraphics[width=2.05\columnwidth]{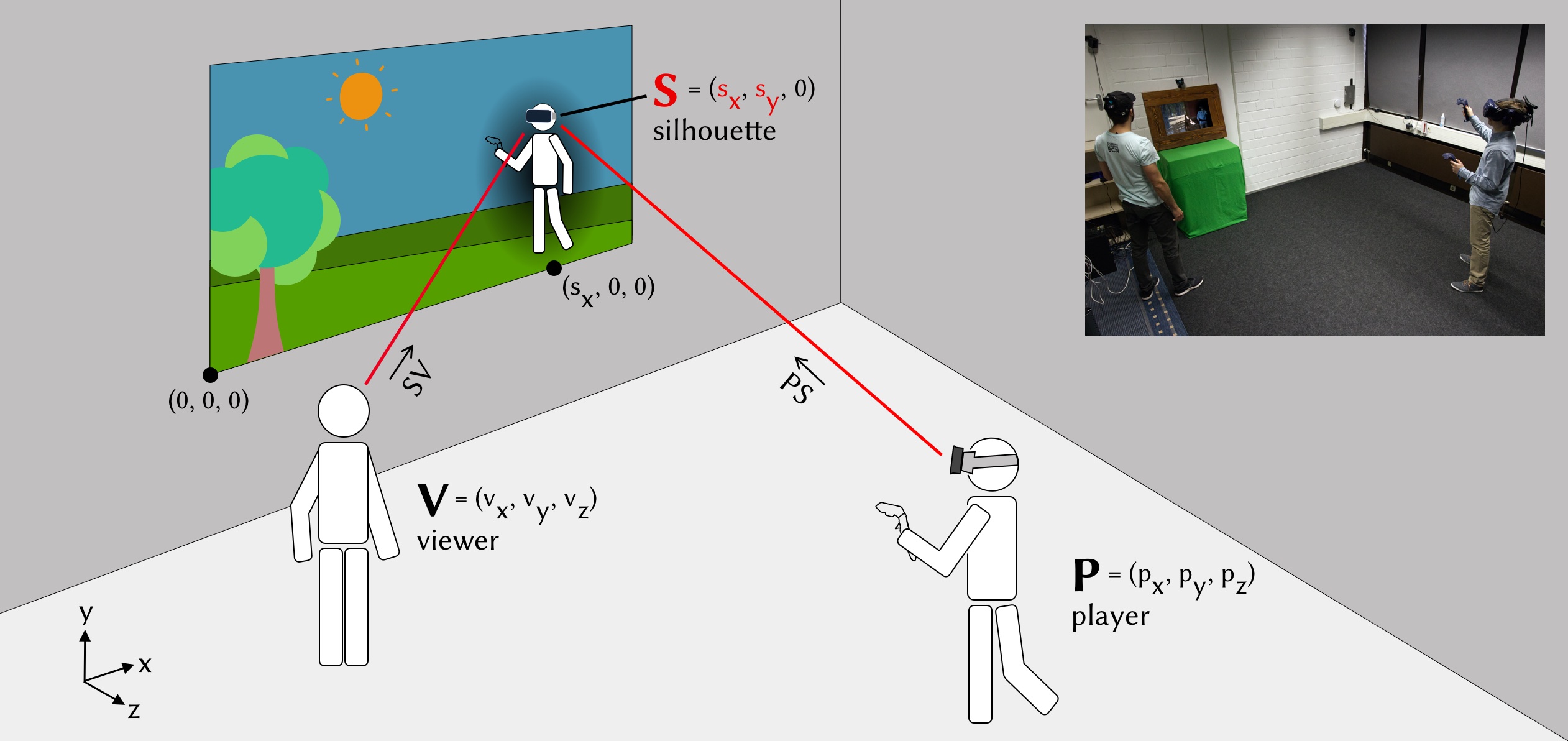}
\caption{ An illustration of the proposed setup. The viewer looks at the screen and sees the player's reflection at the correct position in the game world. To achieve this, we render a silhouette-shaped overlay at this location. The dark silhouette triggers the reflective property of the one-way mirror in front of the screen, blending the player's reflection into VR.}
\label{fig:canvas}
\end{figure*}

\section{The Interactive Mirror Experience}

VR attracts players with a high degree of immersion that such setups offer. Unfortunately, there is no established way to transfer this feeling of being in the virtual world to the audience---rather than to equip the viewers with a VR HMD, of course~\cite{larsson2001actor}. As a result, watching someone playing in VR is not very entertaining. First, we cannot influence what we see, as the view is either predefined (green screen setup) or controlled by the player (first-person view). Second, it is hard for us to set the real-world motions and interactions of the nearby player into the virtual context (first-person view).

This section describes an alternative solution that significantly increases the interactivity on the viewer's side and allows us to see the real player in the virtual world without a green screen capturing setup. By avoiding the impediments mentioned before, our goal is to create a worthy and enjoyable watching experience for VR games.

\subsection{Overview}

We recommend the readers to inspect \FG{fig:canvas} to get a first impression of our technique. In short, we rely upon a one-sided mirror in front of a screen to allow the viewer to see both the virtual world and the player's reflection at the correct virtual position. Hence, we can subdivide our approach into two components: an \textbf{interactive frustum} for the observer and a \textbf{reflection-based visualization} of the player in the virtual world. 

The interactive frustum renders the game scene based on the observer's position relative to the screen. Thus, stepping closer or moving to the side allows the viewer to change the perspective and witness a game event that would remain unnoticed otherwise. For instance, by repositioning, the viewer might detect an enemy approaching from the side. The viewer can even inform or warn the player if such an interaction is desired. Hence, the screen is rather deemed a window into the virtual world than a static display.

The concept of reflection-based visualization allows us to display the player's physical body in the virtual environment. More precisely, we see the reflection of the player at the correct position in the game world. Therefore, we place a one-sided mirror in front of the screen. Such semi-transparent mirrors are translucent in case of a bright background, i.e., when the screen shows the game content. However, if the screen is turned off (or black), the installation behaves like an ordinary mirror. Hence, we render a black overlay in the screen area where the observer would see the player's reflection. Seeing this reflection ``inside'' the virtual world allows us to perceive the player's actions in the gaming context, which gets us closer to this immersive gaming experience.

The provided explanations together make up a top-level view of our approach. In order to become operational, we need to solve the particular problems associated with the described components: First, we need to calculate and render the view frustum depending on the observer's point of view. Second, we have to compute the position and size of the black overlay based on player and observer locations relative to the mirror. The next sections tackle these two challenges and contribute the respective solutions.

\subsection{Interactive Frustum}
The first crucial part of achieving an interactive, authentic experience is the view-dependent scene rendering on the screen behind the mirror. The viewer must perceive the shown image as a correct reflection of the virtual environment. Just like watching through a real mirror, the frustum is determined by the viewer's perspective and changes depending on the viewer's movements.

Our solution follows the established mirror implementations used in computer graphics literature~\cite{fernando2003cg}. However, we do not apply the final texture to an object within the environment. Instead, we render the texture on the display behind the mirror. This difference requires modifications to the usual approach. The complete implementation consists of three consecutive steps:

\begin{enumerate}[leftmargin=1cm]
\itemsep0em
    \item calculate the mirrored perspective
    \item render to texture
    \item display texture to the screen
\end{enumerate}

In the first step, we calculate the matrices to render a correct image of the reflection. Therefore, we place the scene's camera at the viewer's position. The resulting model-view matrix is multiplied with the mirror's reflection matrix to flip the scene. Next, the camera renders to an intermediate texture. In the final step, we display the texture on the whole screen using a blit operation with the correct texture coordinates calculated from the viewer's position.


\subsection{Reflection-Based Visualization}

To bend the player's reflection into the virtual world, we render a black overlay at the estimated screen location. Let $S = (s_x,s_y,s_z)$ denote this silhouette, as depicted in \FG{fig:canvas}. We define the coordinate origin to be in the bottom left corner of the mirror, thus setting $s_z = 0$. Note that we can easily enforce this restriction by translating the mirror.

$S$ depends on the player's position $P = (p_x,p_y,p_z)$ and the viewer's position $V = (v_x,v_y,v_z)$. We break this 3D problem down into two 2D cases, as $s_x$ are $s_y$ independent of one another. In the following paragraphs, we derive the solution for $s_x$, which applies analogously to $s_y$. 

We confine our analysis to the $xz$-plane with $S = (s_x,s_z)$, $P = (p_x,p_z)$ , and $V = (v_x,v_z)$. The normal of the mirror is given by the vector $n = (0,1)$. By the law of reflection~\cite{lekner2013theory}, the angles between the incoming vector $\vv{PS}$ and the outgoing vector $\vv{SV}$ have to match. By definition of the dot product~\cite{riley2006mathematical}, the following equation holds:

\begin{equation} \label{eq:dot}
    \frac{(V - S)}{|V - S|} \cdot n = \frac{(P - S)}{|P - S|} \cdot n.
\end{equation}

By inserting $V,S,P$, and $n$ into \EQ{eq:dot}, we obtain:

\begin{align*}
    \frac{1}{\sqrt{(v_x - {s_x})^2 + {v_z}^2}}&\left(\begin{array}{c} v_x - {s_x}  \\  v_z \end{array}\right) \cdot \left(\begin{array}{c} 0  \\  1 \end{array}\right) \\
    &= \frac{1}{\sqrt{(p_x - {s_x})^2 + {p_z}^2}}\left(\begin{array}{c} p_x - {s_x}  \\  p_z \end{array}\right) \cdot \left(\begin{array}{c} 0  \\  1 \end{array}\right)
\\
\end{align*}

Simplifying the equation above yields
\begin{equation} \label{eq:dotsimplified}
    \frac{v_z}{\sqrt{(v_x - {s_x})^2 + {v_z}^2}} = \frac{p_z}{\sqrt{(p_x - {s_x})^2 + {p_z}^2}}
\end{equation}

\phantom{empty line}

We invert and square \EQ{eq:dotsimplified} to obtain

\begin{equation} \label{eq:dotinvert}
    \frac{(v_x - {s_x})^2+{v_z}^2}{{v_z}^2} = \frac{(p_x-{s_x})^2+{p_z}^2}{{p_z}^2}
\end{equation}

\phantom{empty line}
Further simplifying \EQ{eq:dotinvert} yields
\begin{equation} \label{eq:quadratic}
    {s_x}^2({p_z}^2-{v_z}^2)+2{s_x}(p_x{v_z}^2-v_x{p_z}^2)+{v_x}^2{p_z}^2-{p_x}^2{v_z}^2 = 0
\end{equation}

We have to differentiate between two cases. If $p_z$ equals $v_z$, Equation \ref{eq:quadratic} is simplified to
\begin{equation} \label{eq:dotinvertsimpl}
    2x(p_x{v_y}^2-v_x{p_y}^2)+{v_x}^2{p_y}^2-{p_x}^2{v_y}^2 = 0
\end{equation}

We solve Equation~\ref{eq:dotinvertsimpl} for ${s_x}$ and get
\begin{equation*}
    {s_x} = \frac{{p_x}^2{v_z}^2-{v_x}^2{p_z}^2}{2(p_x{v_z}^2-v_x{p_z}^2)}
\end{equation*}

If $p_z$ is not equal to $v_z$, we apply the reduced quadratic formula to \EQ{eq:quadratic} and obtain
\begin{equation*}
    s_{x_{1,2}} = -\frac{p_xv_z^2-v_xp_z^2}{p_z^2-v_z^2}\pm \sqrt{\left(\frac{p_xv_z^2-v_xp_z^2}{p_z^2-v_z^2}\right)^2-\frac{v_x^2p_z^2-p_x^2v_z^2}{p_z^2-v_z^2}}
\end{equation*}

The desired solution $s_{x}$ is the one between $p_{x}$ and $v_{x}$, i.e., it fulfills the following requirement:
\begin{align*}
min(v_x,p_x) \leq s_{x} \leq max(v_x,p_x).
\end{align*}


As mentioned before, we have to repeat this computation for the $yz$-plane to get the vertical extent of the silhouette. To obtain the top and bottom coordinates of $S$, we use the HMD position and the location of the player's feet as $p_y$, respectively.

Furthermore, we have to decide on the overall appearance of the silhouette. The possibilities range from simple blobs to realistic 1:1 silhouettes. We postpone this discussion until \SC{sec:eval}, where we outline the first impressions from such different silhouette representations.


\subsection{Hardware Prototype}

\begin{figure}[b]
\centering
\includegraphics[width=1.0\columnwidth]{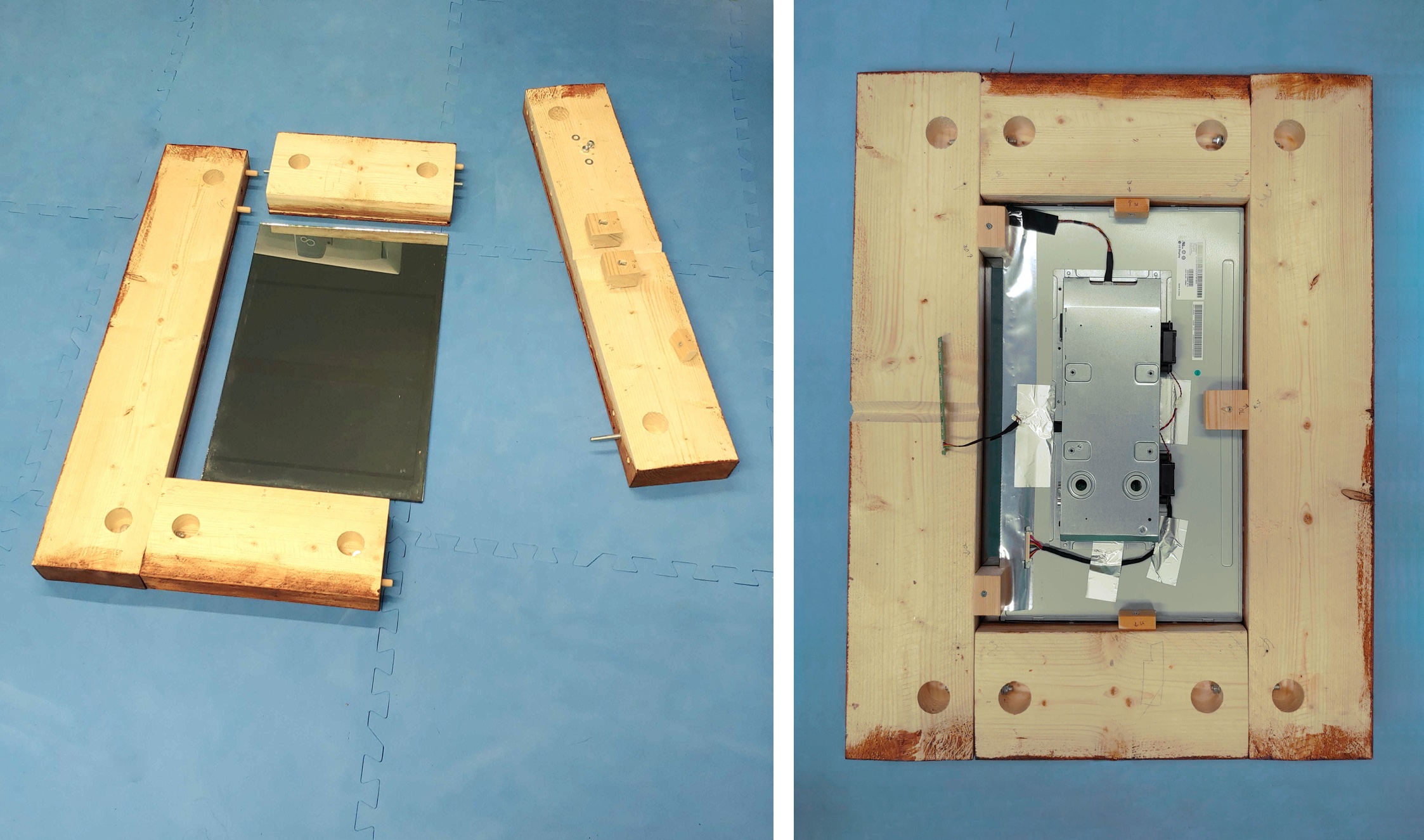}
\caption{For our prototype, we placed a one-way mirror into a  frame and installed a 24-inch display at the back.}
\label{fig:construction}
\end{figure}

We manufactured a small-scale prototype to validate our approach and to have a testbed for qualitative evaluation. In a first step, we carpentered a wooden frame to hold the one-sided mirror, as shown in \FG{fig:construction}. We then disassembled a 24-inch monitor and placed its screen behind the mirror. Such installations are often referred to as ``smart mirrors'', and detailed assembly instructions can be obtained on DIY platforms such as instructables~\cite{instructables}.

To obtain the viewer's position, we mounted a Vive Tracker~\cite{vive} on the backside of a cap (cf. \FG{fig:teaser}). Depending on particular requirements, we suggest alternative solutions, such as marker-less single-camera pose estimation~\cite{cao2018openpose}. Additionally, we attached a tracker on top of the mirror to obtain its current location. This optional add-in increases the overall portability of our setup, as it allows for repositioning the mirror at any time.



\section{Evaluation}
\label{sec:eval}

To our knowledge, the proposed approach is the first of its kind. We therefore administered an evaluation to gather first impressions of how potential viewers would perceive such technology. We were particularly interested in the reception of the interactive view frustum, the refection blending, and the overall impression left by our concept.

\subsection{Study Procedure}

Due to the coronavirus outbreak at the moment of research, we opted for a small-scale qualitative study to gather as many insights as possible from a limited number of participants. The survey took place in a large VR lab at the university to guarantee a sufficient distance between the participant (i.e., the viewer) and the player (i.e., the examiner). After being informed about the study procedure and signing an informed consent, participants completed a questionnaire about their demographic data, gaming habits, and prior VR experiences. Then, the examiner introduced the participants to the operation of our setup (cf. \FG{fig:canvas}). 

\begin{figure}[h!]
\centering
\includegraphics[width=1.0\columnwidth]{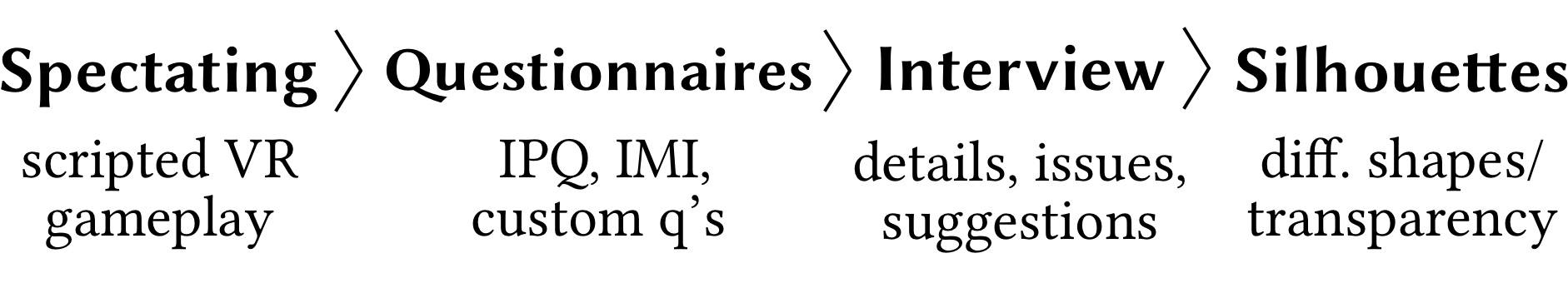}
\caption{An overview of the study procedure.}
\label{fig:procedure}
\end{figure}

The remainder of the survey consisted of four phases as outlined in \FG{fig:procedure}: Firstly, the examiner outlined and played the VR testbed game described below. During this gaming session, the participants acted as viewers and could freely move around the room to adjust their view frustum. Secondly, we administered the enjoyment subscale of the Intrinsic Motivation Inventory (IMI)~\cite{ryan2000self} to assess the general enjoyment resulting from our approach, as well as the Igroup Presence Questionnaire (IPQ)~\cite{Schubert.2003, Schubert.2018} to measure the feelings of presence. In particular, we utilized the IPQ subscales spatial presence, involvement, experienced realism, and general presence. We also posed some custom questions (abbreviated as CQ, cf. \TA{tab:custom}) to assess how participants evaluated specific parts of the installation, such as the view frustum and the player reflection. All administered questionnaire items had to be rated on a unipolar scale ranging from 0 to 6 (``completely disagree'' to ``completely agree''). For IMI, we kept the original 7-point scale. The original IPQ employs a 7-point scale and suggests ratings from -3 to +3. We wanted to keep our questions uniform for the participants and shifted the scale to 0-6. Thirdly, the examiner conducted a semi-structured interview aimed at the identification of particular strengths and weaknesses of our viewing technology. For instance, we asked, ``How did you perceive the interactions of the player with the surrounding objects?'' (see supplementary material for the complete list of questions). Fourthly and finally, the participants revisited the setup, where the examiner demonstrated our silhouette variations (cf. \FG{fig:silhouette}) in a randomized order. We presented these alternatives to determine a possible sweet spot for future applications. In particular, we varied the width and transparency of the blob. Furthermore, we offered a silhouette that mimics the player's arm movements by repeating our silhouetting algorithm for the two tracked controller coordinates. We assessed each silhouette's perception by multiple questions, e.g., ``Do you perceive the player as part of the virtual environment?'' (cf. supplementary material). In the end, we asked the participants for final comments and debriefed them.

\begin{figure}[t]
\centering
\includegraphics[width=1.0\columnwidth]{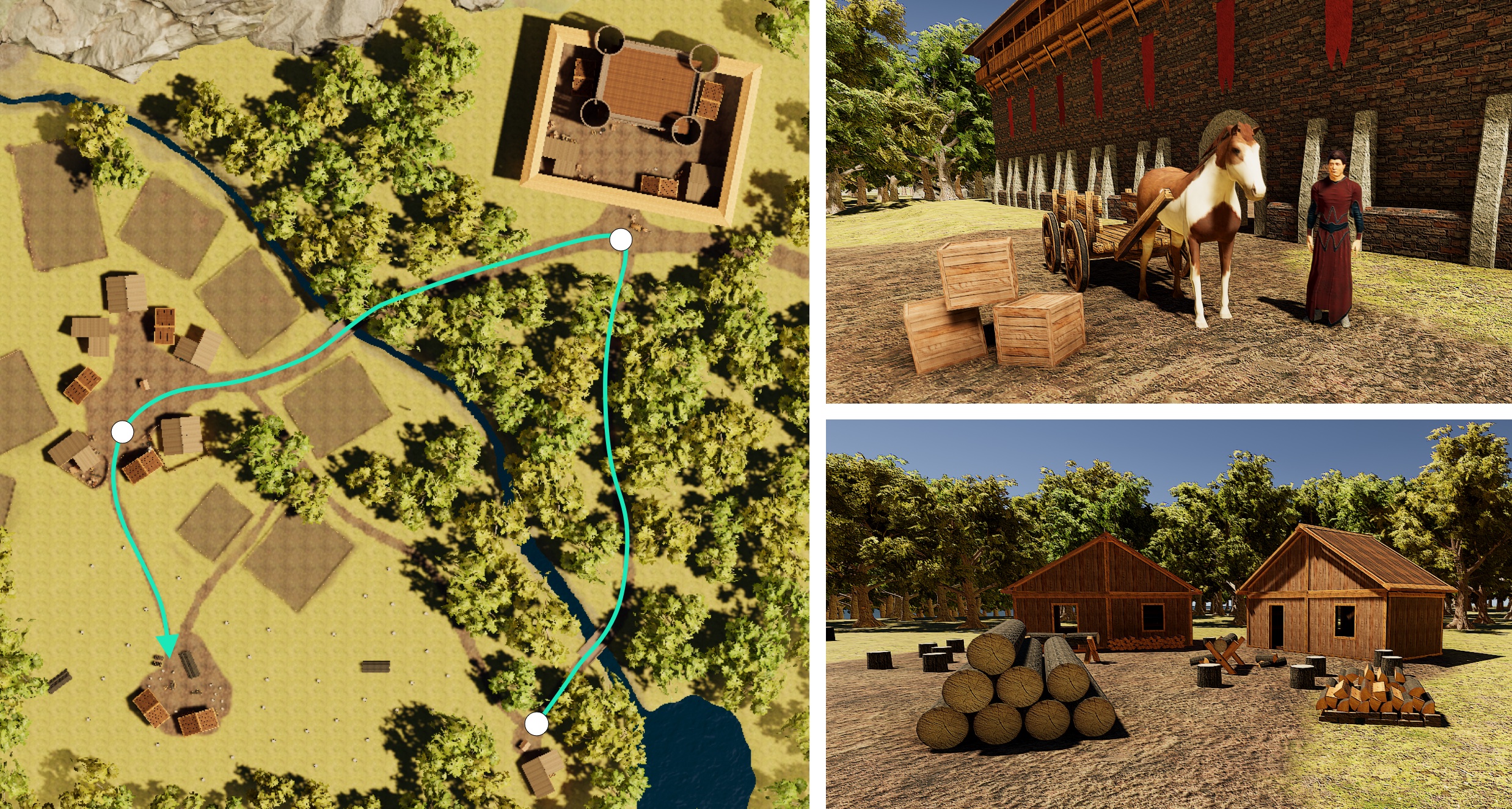}
\caption{The testbed adventure game. The left image depicts the path taken by the player, i.e., the examiner. Two game quests were interactions, such as sorting crates or collecting wood (right). The two other quests focused on traveling.}
\label{fig:game}
\end{figure}

\begin{figure}[b]
\centering
\includegraphics[width=1.0\columnwidth]{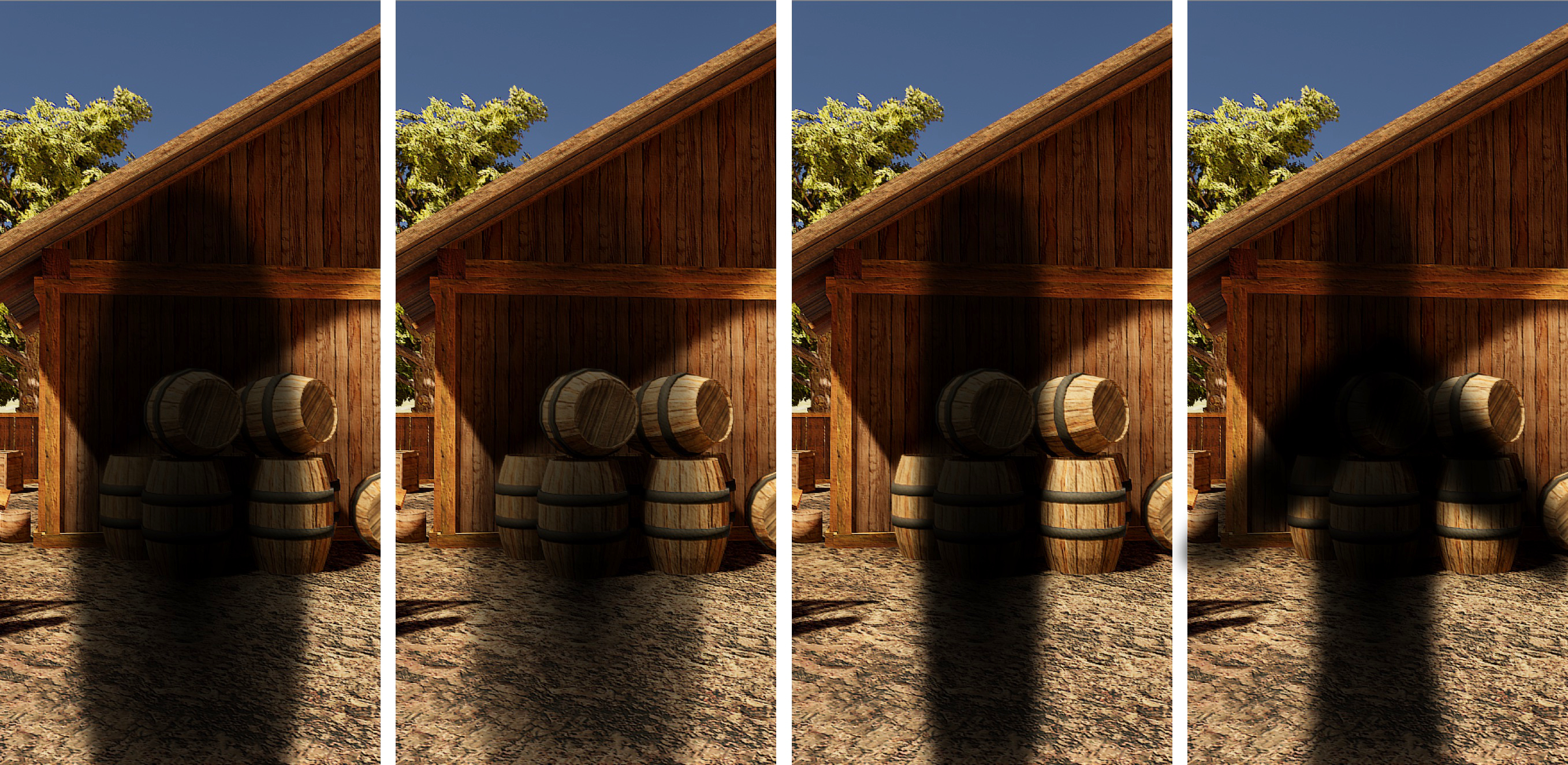}
\caption{We demonstrated a variety of silhouettes at the end of the study. From left to right: default oval (used during the gaming session), increased transparency, narrow oval, body-shaped silhouette with tracked arms.}
\label{fig:silhouette}
\end{figure}

\begin{table*}[]
\caption{Mean scores (range 0-6) and standard deviations of the custom questions (CQ).}
  \label{tab:custom}
  \begin{tabular}{llr}
    \toprule


    CQ1 & I understood what the player was doing in VR at any point during operation. & 5.11 (0.78) \\\addlinespace

    CQ2 & I could see well how the player interacted with the in-game objects. & 5.33 (0.71) \\\addlinespace 

    CQ3 & I did not feel like I missed any essential game events. & 3.78 (1.64) \\\addlinespace

    CQ4 & I was able to control my perspective on the game world freely. & 4.11 (1.27) \\\addlinespace

    CQ5 & I could not see everything I wanted in the game world. & 4.56 (1.59) \\\addlinespace

    CQ6 & I was in the same world as the player. & 4.00 (1.12) \\\addlinespace

    CQ7 & I saw the virtual world as if I was there. & 2.67 (0.87) \\\addlinespace

    CQ8 & I enjoyed my ability to explore the virtual world via repositioning and head movements.  & 5.56 (0.53) \\\addlinespace
    CQ9 & It was fascinating to see the virtual world through the mirror. & 5.22 (0.44) \\\addlinespace

    CQ10 & I would have liked to interact with the player. & 4.67 (1.66) \\\addlinespace
    CQ11 & I was part of the game. & 2.22 (1.20) \\\addlinespace
    CQ12 & I felt passive throughout the gaming session. & 5.00 (1.32) \\\addlinespace

    CQ13 & I would have preferred a larger screen. & 5.78 (0.44) \\\addlinespace
    CQ14 & The view frustum was too small for me. & 5.22 (0.83) \\\addlinespace
   
\end{tabular}
\end{table*}
\subsubsection{Testbed Game.} We used a medieval 3D adventure game designed by Krekhov et al.~\cite{gullivr} as our VR scenario (cf. \FG{fig:game}). In this game, the player takes up the role of an herbalist’s apprentice and solves four different quests. The first and third quest focus on object interaction, i.e., relocating or arranging items. The second and the fourth quest require the exploration of the surrounding via point and teleport locomotion~\cite{bozgeyikli2016point}. Hence, the participants experienced local, player-oriented interactions as well as long-distance travel in an open world. 

The overall duration of the gaming session was approximately 15 minutes. To guarantee similar conditions for all participants, we scripted the examiner's gameplay. We neither restricted nor encouraged the communication between the participants and the examiner. However, the examiner was advised to ignore major requests (e.g., going to a different location) to prevent significant alterations in experienced gameplay.

\subsection{Results and Discussion} 

Because of the pandemic situation, we limited the number of participants to nine persons (three female, six male) with a mean age of 27.8 years (\textit{SD}~=~7.66). Four of them were students, and the others were employees. All participants reported playing digital games regularly, and eight of them had prior experiences with VR HMDs. An excerpt of our custom questions is outlined in \TA{tab:custom}, and the outcomes of the IMI and the IPQ questionnaires can be found in \TA{tab:default}. To evaluate IMI, we followed the respective guidelines~\cite{ryan2000self}, i.e., we first computed the mean of all seven items for each participant and used this data for further computations. 

The interviews and the discussion on the silhouette shapes were audio-recorded with the permission of the participants. We applied the six-phase process of reflexive thematic analysis by Braun et al.~\cite{thematic} to analyze the interview data, following a deductive approach. As a result, we identified themes such as ``player locomotion'' or ``reflection/visibility.'' In the following, we group all obtained results and insights into respective thematic areas.

\begin{table}[b]
  \caption{Mean scores (range 0-6) and standard deviations of the IMI and IPQ subscales.}
  \label{tab:default}
  \begin{tabular}{llc}
    \toprule
    \textbf{IMI}
    & Enjoyment/Interest & 4.51 (0.53) \\
    \noalign{\vskip 2mm}   
    \textbf{IPQ} 
    & General &  3.22 (1.20) \\
    & Spatial presence  &  3.56 (0.92)\\ 
    & Involvement  &  2.94 (1.29)\\
    & Experienced Realism  &  2.53 (0.48)\\
  \bottomrule
\end{tabular}
\end{table}

\subsubsection{Overall Concept.} The first questions of our semi-structured interview targeted at the overall impression left by our setup. Seven participants explicitly told us that they were positively surprised of the overall experience, as they \textit{``didn't think that it might be possible to exploit reflections in such a way'' \textbf{(P4)}} and described it as an \textit{``unusual but refreshing solution'' \textbf{(P7)}}. The high values for CQ8 (\textit{M}~=~5.56, \textit{SD}~=~0.53) and CQ9 (\textit{M}~=~5.22, \textit{SD}~=~0.44) expose the overall excitement of the participants regarding the possibility to control their view frustum via repositioning. One participant was particularly impressed by the performance: \textit{``I expected some lags and delays when I heard about the dynamic viewport. But then, it was smooth, without flicker or lag, even during fast head movements'' \textbf{(P4)}}.

This positive overall impression is supported by the high values of the IMI enjoyment/interest subscale (\textit{M}~=~4.51, \textit{SD}~=~0.53). This outcome is even more conclusive under the circumstance that the participants did not play the game themselves and were only in an observer role. This initial appreciation could be due to the innovative character of our setup. We suppose that this novelty effect might wear off after a while. Nevertheless, the gathered insights support our claim that the proposed technique leads to an enjoyable experience and is worth further explorations.



\subsubsection{Perception of the Player's Reflection.} Our motivation behind the reflection utilization was to allow the viewer to see the player inside the virtual world. We suggest that such a third-person view enhances the comprehensibility of player's actions due to the additional context provided to the audience~\cite{reeves2005designing}. This assumption is supported by the high values for CQ1 (\textit{M}~=~5.11, \textit{SD}~=~0.78) and CQ2 (\textit{M}~=~5.33, \textit{SD}~=~0.71). These outcomes are in line with the insights from our interviews, where participants reported to \textit{``understand what he [the examiner] was doing with the objects in front of him'' \textbf{(P1)}}. The participants also mentioned that \textit{``especially the moments when the player engaged with an NPC were most interesting to watch'' \textbf{(P8)}}. Overall, the interaction aspect was perceived as very immersive: \textit{``when he [the examiner] threw a crate, or the scroll it felt like these objects will fly at me'' \textbf{(P5)}}.

However, the fact that the viewer can see both the real player and the reflection is also a potential source of confusion. For example, one participant stated that \textit{``in situations where the player physically approached me from behind, I had a strong need to turn around'' \textbf{(P1)}}. Another participant, in contrast, \textit{``did not perceive the real player anymore and only looked at the VR mirror'' \textbf{(P3)}}. Also, some participants pointed out the depth ordering issue, as our silhouette computation is currently limited to the screen space: \textit{``sometimes it was hard to say whether an item was in front or behind the player, particularly when it was covered by the dark blob'' \textbf{(P6)}}. To fix the depth order, we need to include the depth coordinate $s_z$ in our future implementation in the same computational way as we did with $s_x$ and $s_y$.

\subsubsection{Shape of the Silhouette.}

We can not derive a general trend regarding the optimal appearance of the silhouette, as we received controversial opinions on this topic. What we can say is that the complex silhouette that included the player's arms was disliked by all participants. The reasons were manifold, e.g., four participants complained that \textit{``the dark arm notches were too distractive and occluded too much'' \textbf{(P2)}}. Even worse, two participants \textit{``could not see the reflected arms properly within the dark area'' \textbf{(P7)}}. One participant rated this silhouette as \textit{``unrealistic, resembling more this skydancer or tube man thing'' \textbf{(P1)}}.

Similarly, there was no common agreement regarding the width of the silhouette. One participant strongly disliked the narrow representation because \textit{``it was hard to see the player'' \textbf{(P3)}}. Otherwise, participants did not express any tendency toward the wider or more narrow shape. Regarding transparency, only one participant preferred a highly transparent or no silhouette at all. Six of the remaining participants requested a transparency level depending on the virtual surrounding, asking us to \textit{``make the silhouette darker when the player is in a bright and feature-rich environment'' \textbf{(P9)}}. In contrast, dark areas barely require a silhouette to see the player's reflection.

\subsubsection{Immersiveness of the Interactive Frustum.} We designed the dynamic frustum to lower the HMD-induced barrier between the player and the viewer. CQ6 (\textit{M}~=~4.00, \textit{SD}~=~1.12) supports our assumption, as participants felt like they are in the same world as the player. We argue that this world is the physical realm and not the virtual surrounding, as indicated by the low values for CQ7 (\textit{M}~=~2.67, \textit{SD}~=~0.87). In other words, the frustum brings the player into the living room, rather than the viewer into the virtual environment. This explanation also aligns with the values for the general (\textit{M}~=~3.22, \textit{SD}~=~1.20) and the spatial presence (\textit{M}~=~3.56, \textit{SD}~=~0.92) subscales of the IPQ.

As indicated by CQ4 (\textit{M}~=~4.11, \textit{SD}~=~1.27), the participants had no trouble to control the view frustum according to their wishes and enjoyed such a possibility to explore the virtual world (CQ8; \textit{M}~=~5.56, \textit{SD}~=~0.53). Three participants explicitly described the experience as \textit{``if I would look out of a window and see the game world'' \textbf{(P8)}}. More importantly, they barely missed out on any important events (CQ3; \textit{M}~=~3.78, \textit{SD}~=~1.27). However, the additional degree of freedom provided by the frustum is hindering during player locomotion. Seven participants mentioned a loss of orientation during the teleportation procedure. The point and click teleport was described as \textit{``instant and unpredictable jump that I could not control and needed a while to find some reference points in the scene to know where we [the participant and the examiner] arrived'' \textbf{(P2)}}. Two participants disclosed that they \textit{``tried to reposition the view such that the teleport target and the player are both visible'' \textbf{(P5)}}. Hence, we assume that our setup has a notable drawback for games that involve a large amount of player relocations, as the viewers need a certain amount of time to reorient themselves in the virtual world.

\subsubsection{Limited Screen/Mirror Size.} The major point of criticism was the size of the view frustum. Nearly all participants requested a bigger screen/mirror (CQ13; \textit{M}~=~5.78, \textit{SD}~=~0.44) and remarked that the viewport was too limited (CQ14; \textit{M}~=~5.22, \textit{SD}~=~0.83). Accordingly, the participants could not see everything that they wanted to see (CQ5; \textit{M}~=~4.56, \textit{SD}~=~1.59). Three participants told us that \textit{``sometimes, it was hard to find a proper angle that would reveal the player and the area of interest in the game world'' \textbf{(P2)}}. Similarly, one participant explained: \textit{``it looks best when the player entirely fits into the screen from head to foot, but this is hardly possible because of the small screen size'' \textbf{(P9)}}. This screen size issue was also reflected in the viewers' behavior during the gaming session---most viewers stood close to the mirror, which results in a smaller player silhouette and an increased field of view. Thus, we must conclude that a 24-inch screen is definitively too small and suggest to utilize a minimum of 50 inches in the future.

\subsubsection{Missing Interaction with the Player.} The lowered barrier between the player and the viewer, combined with the feeling of being in the same world, naturally evokes a need for interactivity. Hence, we postulate that our setup is especially interesting for collocated gaming purposes~\cite{cox2016public, kappen2014engaged, wehbe2015towards}. As indicated by CQ11 (\textit{M}~=~2.22, \textit{SD}~=~1.20), the participants did not perceive themselves as a part of the game and felt rather passive (CQ12; \textit{M}~=~5.00, \textit{SD}~=~1.32). Instead, they would have liked to interact with the player (CQ10; \textit{M}~=~4.67, \textit{SD}~=~1.66).

We see this input as an important aspect of future research, as our installation could lead to novel playful experiences that transform the viewer from a passive spectator~\cite{downs2015differentiated} to an equal co-player. For instance, one participant proposed to \textit{``include game objects that are visible only for the observer'' \textbf{(P4)}} or to \textit{``turn the watcher into an early warning system for enemies sneaking up from behind'' \textbf{(P4)}}. Such an asymmetric information distribution would upgrade the spectators from ``witnesses'' of the story to actors or even ``heroes''~\cite{nicolae2018spectator}. Another participant mentioned mixed reality sports games as a possible application area: \textit{``it could be fun to play some sort of tennis, me in front of the mirror and my opponent in VR'' \textbf{(P2)}}. For this purpose, one could blend the real surrounding into VR, as proposed by Hartmann et al.~\cite{hartmannRealityCheck}, or even turn the room into a shared mixed reality environment, as done by Jonas et al.~\cite{jones2014roomalive}.

\section{Limitations}

The outlined setup is the first of its kind. Hence, the method and its evaluation come with weaknesses and limitations that have to be discussed explicitly. This section is dedicated to a critical, retrospective assessment of the presented work, as we aim at providing a comprehensive overview of the pros and cons of our system and paving the way for meaningful future work in this direction.

\subsection{Limitations of the Study}

Due to the restrictions caused by the pandemic, we had to limit our evaluations to a bare minimum. Having only nine participants limits the conclusiveness of the gathered results. For instance, this small sample does not allow us to draw general assumptions on the optimal silhouette shape---instead, we consider such outcomes as preliminary insights. Another issue is that eight out of nine participants were already familiar with HMDs, which induces an initial bias. For instance, inexperienced participants might rate the spectatorship role less engaging, as they would be more interested in a first-hand VR experience. 

Furthermore, the study procedure consisted of multiple steps (cf. \FG{fig:procedure}), and each participant underwent multiple evaluation rounds, e.g., observing the gameplay, participating in an interview, and returning to the game to rate various silhouettes. This complexity could have caused fatigue and bias: for instance, seeing the complete gameplay with one specific silhouette shape could have impacted the final assessment of such shapes.

Our evaluation was limited to an explorative survey and aimed at gathering first impressions left by the technique. Hence, at this moment, we cannot provide meaningful comparisons to other spectatorship techniques. This is an important limitation and needs to be addressed in a follow-up quantitative study after the pandemic. In particular, there is a need for a baseline comparison between our mirror, a default first-person perspective, and different third-person solutions (see future work section).

Another restriction of our study is the utilization of only one testbed game. As we have seen in our case, the appreciation of the technique depends on the game events. Teleportation and long-distance navigation caused disorientation, rendering such events rather unattractive for spectators. In contrast, local interactions, e.g., picking up or moving objects, were perceived as engaging and immersive. Hence, we have to consider a diverse set of games before drawing overarching conclusions on our technique's general applicability.

Regarding the applied measures, two shortcomings have to be mentioned. Firstly, we did not instruct the participants regarding the interaction and communication with the player. Thus, despite the outcomes of CQ10, there were no interaction attempts. We argue that this behavior is due to the formal study atmosphere, amplified by the pandemic influence. Secondly, we focused solely on the viewer's experience and ignored the player in our setup. However, to become established, the technique needs to provide an adequate player experience as well. For instance, the colocated spectator could cause discomfort of the players, as, e.g., outlined by Rogers et al.~\cite{10.1145/3290605.3300644}.

\subsection{Restrictions of the One-Way Mirror Setup}

The initial implementation of the one-way mirror idea comes with inherent strengths and weaknesses and should not be considered an all-round solution. In our version, the interactive frustum is computed based on the viewer's position. Hence, such a setup is currently limited to one observer---an issue for, e.g., public spaces with multiple interested persons.

Furthermore, the technique requires certain modifications to the game (engine) to render two perspectives---one for the player and one for the observer. This overhead also implies increased hardware requirements. We point out that the observer's perspective does not require the same high refresh rates as the VR version. A further hardware-related restriction is a need for additional components, i.e., a one-way mirror and two default HTC Vive Trackers, limiting the out-of-the-box applicability of the method.

Another critical design consideration is the visualization of the player. In our case, we used the player's reflection. Like a green screen or \textit{RoomAlive}~\cite{jones2014roomalive}, the viewer sees the real player, which fits less well into the virtual environment compared to an avatar. This break in the appeal can lead to a decrease in perceived presence. On the other hand, only a few VR games provide a player model. Although some green screen solutions, such as LIV~\cite{LIV}, provide fallback models, we argue that an inappropriate choice of the avatar can hamper the immersion even further. Besides, we suggest that seeing the real player could enhance the bond between player and viewer and provide a refreshing experience compared to default avatars.

\section{Future Work}

The mentioned limitations provide an outline for further research on the one-way mirror idea. In a first step, we suggest an in-depth, quantitative evaluation of our technique to understand how it performs versus established solutions. We plan to conduct a baseline comparison to the following candidates:
\begin{enumerate}[leftmargin=1cm]
\itemsep0em
    \item first-person perspective: the viewer sees the player's perspective; i.e., the default spectator scenario
    \item third-person perspective (player): a green screen is used to inject the player into the virtual environment
    \item third-person perspective (avatar): again, we capture the player's movements, but the player is replaced by an avatar in-game
\end{enumerate}
Depending on the outcomes of (2) and (3), one might also consider adding an avatar mode to our technique. Therefore we have to disable the dark silhouette and render the avatar at the respective position. Furthermore, we suggest assessing the player experience in our setup. We hypothesize a slight discomfort~\cite{10.1145/3290605.3300644,10.1145/2207676.2208347} caused by the invisible, yet colocated spectator. In this case, our technique could benefit from a minimalistic visualization of the viewer from the player's perspective.

One possible follow-up of our setup is its adaptation to a distributed scenario. This is especially viable in the times of the coronavirus pandemic, as it supports social distancing. A remote session in our case would even allow multiple spectators, each equipped with an individual, dynamic view frustum, e.g., a tablet. Note that this variation requires a video capturing device on the player's side, as we cannot rely on the direct reflection anymore.

Our evaluation results suggest that the display size needs to be extended in future hardware iterations. The small screen forced the viewers to stand very close to it, which restricts mobility and limits the idea's overall potential. Another point that can be improved is a more sophisticated attenuation of player teleportation. In our study, the viewers experienced a loss of orientation during fast travel, which could be reduced by supporting VFX/SFX effects that signal upcoming teleportation.

Finally, our study participants expressed a strong need to interact with the player, be it verbally or playfully. We gathered some preliminary suggestions, including sports games and games with asymmetric information distribution. We suggest implementing such testbed scenarios to increase the viability of the presented setup. Ultimately, such games will pave the way to more inclusive, engaging, and active spectatorship experiences.

\section{Conclusion}

Our work presented a novel possibility to watch players in VR. Instead of looking at a static display, the spectator can change the point of view by repositioning and changing the viewing angle. This approach turns a common screen into a view-dependent window, thus significantly increasing the autonomy of the viewer. The setup is completed by a one-way mirror in front of the display. Together with our silhouetting algorithm, this mirror allows the spectator to see the player's reflection at the correct position in the game world, blending the player into the virtual environment. The result is similar to a green screen caption, yet without the capturing overhead and, most importantly, with a spectator-controlled viewport.

Our exploratory study underpins the overall positive appreciation of our technique. The reflection-based player visualization renders object and NPC interactions comprehensible and engaging. The dynamic view frustum enables autonomous exploration of the surroundings and prevents viewers from missing out on interesting game events. Taken together, these two components of our system lower the HMD-induced barrier between viewers and players, evoking the feeling of being in the same world. Hence, our collocated setup provides a solid foundation for novel playful experiences that transform a passive spectator into an engaged, mixed-reality co-player.

\bibliographystyle{ACM-Reference-Format}
\bibliography{mirror}

\end{document}